\documentclass[aps,preprint,showpacs]{revtex4}
\usepackage{amsmath, amssymb} 
\usepackage{graphicx}
\begin{document}
\
\title{A formula for the minimal coordination number of a parallel bundle} 

\author{E.L. Starostin}

\affiliation{Centre for Nonlinear Dynamics, Department of Civil, Environmental \& Geomatic Engineering, 
University College London, Gower Street, London WC1E 6BT, UK\\
Email: e.starostin@ucl.ac.uk; eugene\_starostin@hotmail.com}
\pacs{87.15.-v, 87.16.Ka, 61.46.Fg} 

\begin{abstract}
An exact formula for the minimal coordination numbers of the parallel packed bundle of rods is presented
based on an optimal thickening scenario.
Hexagonal and square lattices are considered.
\end{abstract}

\maketitle

\section{Introduction}

Hexagonal packing of slender objects (filaments or rods) like semiflexible polymers, e.g. DNA~\cite{Livolant89,Durand92,Livolant96}, filamentous actin~\cite{Tilney75}, collagen~\cite{Hulmes79} or nanotubes~\cite{Thess96,Colomer04,Needleman04} provides the most compact way of filling a domain in space.
It maximizes the number of contacts between the rods. 
It has long been known that DNA can condensate to form bundles, in particular, closed toroids when polyvalent cations are added to the solution~\cite{Evdokimov72,Marx83,Boettcher98,Golan99,Hud05}.
Similar shapes are observed for other filaments stabilized by different types of adhesion: by hydrogen bonding, by ionic interactions or by van der Waals forces~\cite{Tang01,Martel99}.
A perfect hexagonally packed bundle must be twistless to provide continuous parallel arrangement of filaments though it can writhe in space~\cite{Starostin06}. On the other hand, the interaction
peculiarities of individual filaments in a parallel bundle may favour them to fill in a non-hexagonal (e.g., square) 2D lattice in cross section.

The main purpose of this paper is to present a compact exact formula for a number of unoccupied bond sites on the surface of a tightly packed parallel bundle. This quantity commonly appears in analysis of stability of DNA toroids~\cite{Bloomfield91,Pereira00,Schnurr02,Cherstvy05,Stukan06,Ishimoto08} and 
it has been computed either approximately~\cite{Bloomfield91,Schnurr02} or by a complicated procedure~\cite{Pereira00,Stukan06} or for particular numbers of strands in the bundle~\cite{Cherstvy05,Ishimoto08}.
It is reasonable to believe that the exact analytical expression will prove to be handy in the future analytical research.

Consider a bundle of parallel packed $N$ rods. In the Euclidean space, each rod may be in continuous contact with at most 6 other rods~\cite{Starostin06a}. 
The total number of binding sites is $6N$ for hexagonal lattice and
$4N$ for square one. Each actual bond involves two sites. Therefore, the number of the binding sites that are exposed to the environment is even and one half of it is called the coordination number~\cite{Schnurr02}.

In a poor solvent, there is a relative energetic advantage for a filament to have a polymer-polymer contact versus being exposed to solvent.
Therefore, we are interested in the tightest bundle when the coordination number is minimal.
The minimal coordination number enters a free energy expression that is used in studies of DNA condensation~\cite{Bloomfield91,Pereira00,Schnurr02,Cherstvy05,Stukan06,Ishimoto08}. Its behaviour as a function of number
of filaments affects stability of various conformations.

Since the rods are parallel, all orthogonal cross sections of the bundle are congruent and we can treat the problem reduced to two dimensions, i.e. we consider the set of discs that appear in an orthogonal cross section of the bundle.
The minimal coordination number equals
\begin{equation}
\alpha_N=\left\lceil\sqrt{12 N - 3}\right\rceil,
\label{eq:coordnmb}
\end{equation}
for the hexagonal lattice and
\begin{equation}
\beta_N=\left\lceil2\sqrt{N}\right\rceil,
\label{eq:coordnmb_sq}
\end{equation}
for the square lattice. In Eqs.~(\ref{eq:coordnmb}),(\ref{eq:coordnmb_sq}),
$\lceil\cdot\rceil$ stands for the ceiling function: for $x \in \mathbb{R}$, $\lceil x\rceil$ is defined as the smallest integer greater than or equal to $x$.
In particular, $\alpha_1 = 3$, $\alpha_2 = 5$, $\alpha_3 = 6,$ and so on and
$\beta_1 = 2$, $\beta_2 = 3$, $\beta_3 = 4,$ and so on.

Equation~(\ref{eq:coordnmb}) gives exactly the same values (up to factor of 2) as a rather overcomplicated algorithm that was suggested
in Ref.~\cite{Pereira00}, Eq.~(3), and was used recently in Ref.~\cite{Stukan06}.
An approximate expression was presented earlier for the average number of interactions which equals $6-2\alpha_N/N$ in our terms (see
Eq.~(10) in Ref.~\cite{Bloomfield91}). That approximate formula may be used to compute the coordination number only for small
number of members, up to $N \lesssim 30$. Actually, Eq.~(10) describes a nonmonotonic function which contradicts the physical sense.
Note also that the graph in Fig.~3 in the cited article which is claimed to show the function from Eq.~(10), in reality,
represents a different non-specified function. 

For a particular case of the bundles with a complete hexagonal cross section,
one can find an exact expression for the average number of interactions in Ref.~\cite{Cherstvy05} (Eq.~(4)) which agrees with Eq.~(\ref{eq:coordnmb}) if
$N=3n^2+3n+1, n = 1,2,\ldots$. It was suggested to use the formula of the number of bonds computed for complete hexagons to approximate
the optimal configurations for artbitrary number of members~\cite{Ishimoto08}. 
The limiting coordination number $\alpha_\infty = 2 \sqrt{3N}$ proposed in Ref.~\cite{Schnurr02} approximates
Eq.~(\ref{eq:coordnmb}) for a hexagonal bundle of $N$ filaments when $N \to \infty$.
No proof was given on optimality of conformations considered in the above mentioned works to compute the minimal coordination number.

\section{Proof of Eq.~(\ref{eq:coordnmb})}

\begin{figure}[htbp]                        
\begin{center}
\includegraphics[width=4cm]{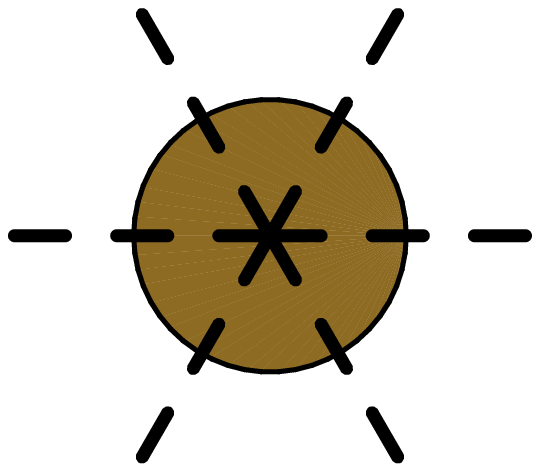} \quad \includegraphics[width=4cm]{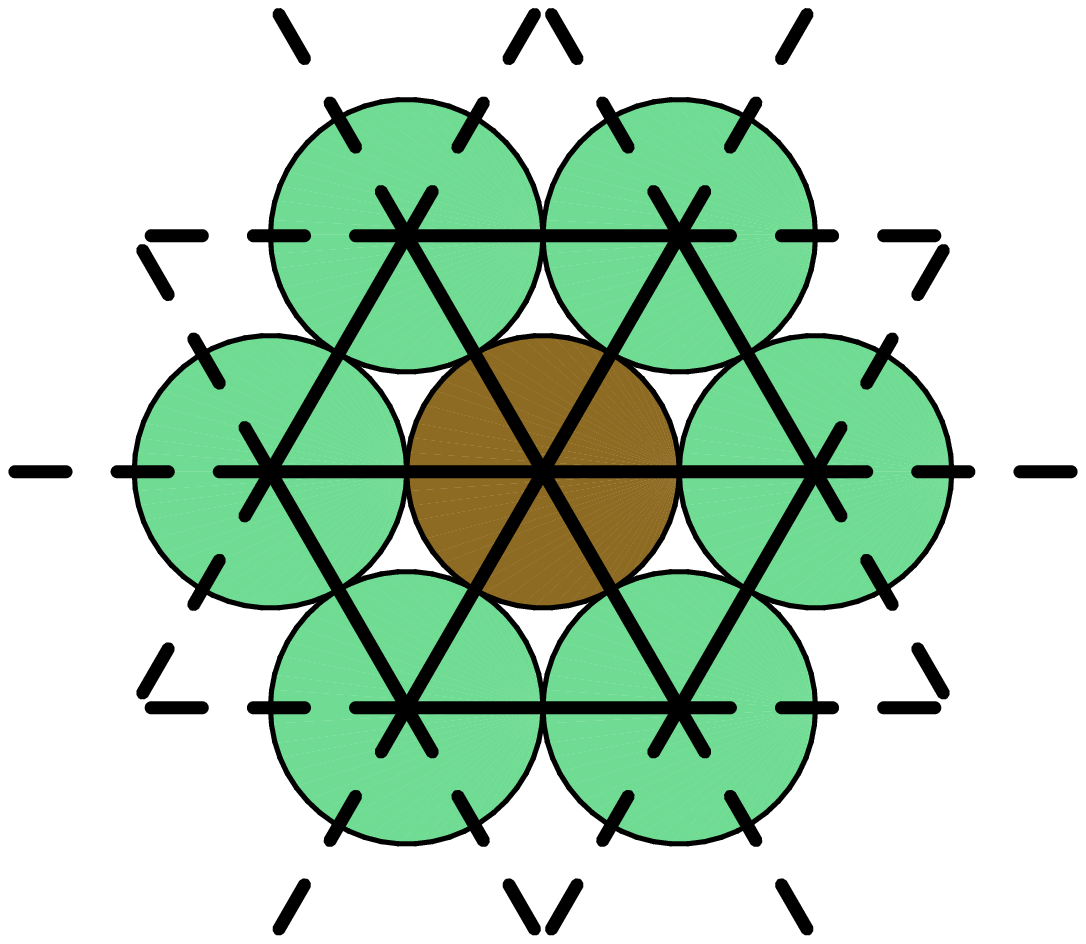} \quad \includegraphics[width=4cm]{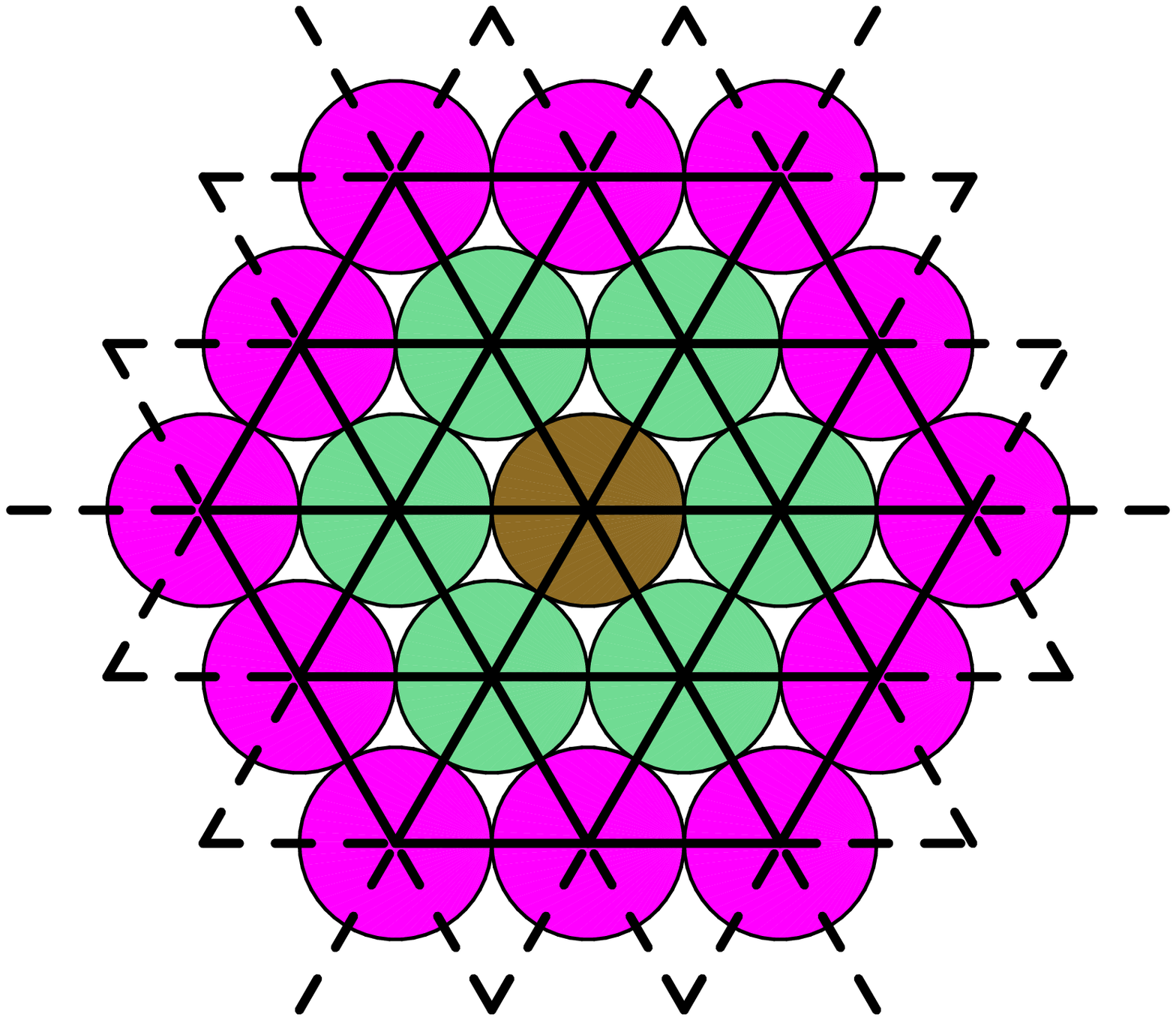}
\end{center}
\caption{Complete shells of generation $n=1,2,3$. Here and in all other figures, bonds are shown as black sticks, free potential bonds are dashed.}
\label{f.1}
\end{figure}

1. We show that Eq.~(\ref{eq:coordnmb}) is valid for complete shells, i.e. for $N = M(n)=3n^2-3n+1, n = 1,2,\ldots$ (Fig.~\ref{f.1}). Clearly, $\alpha_1=3$ for $n=1$. For $n = 2,3,\ldots$, there are 6 corner cells, each with 3 open bonds and $6(n-2)$ side cells, each with 2 open bonds. 
In total, $6\times 3+(6n-12) \times 2 = 2(9 + 6n -12) = 12n-6 \equiv 2 \alpha_{M(n)}$~\cite{Pereira00}.

\begin{figure}[htbp]                        
\begin{center}
\includegraphics[width=3.90cm]{shell1.eps} \includegraphics[width=3.90cm]{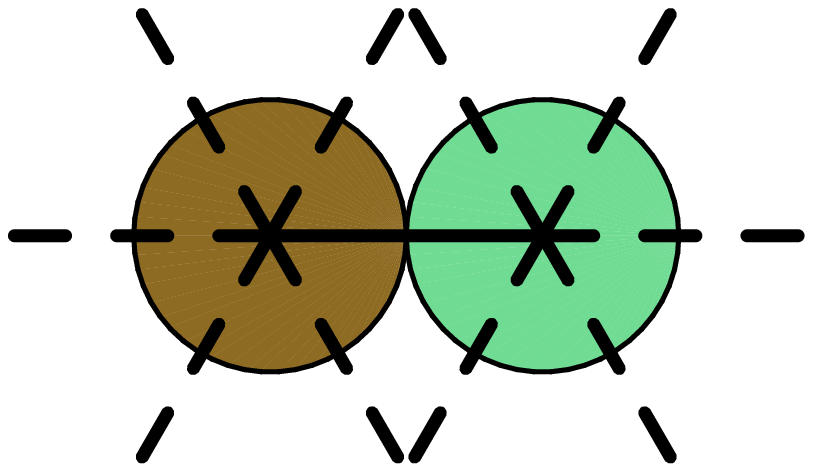}  \includegraphics[width=3.90cm]{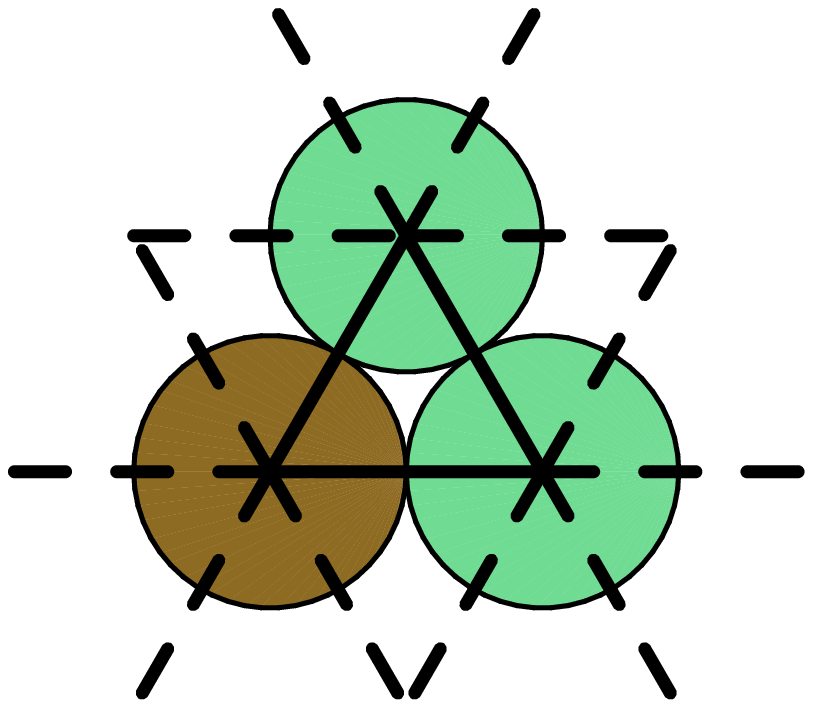} \includegraphics[width=3.90cm]{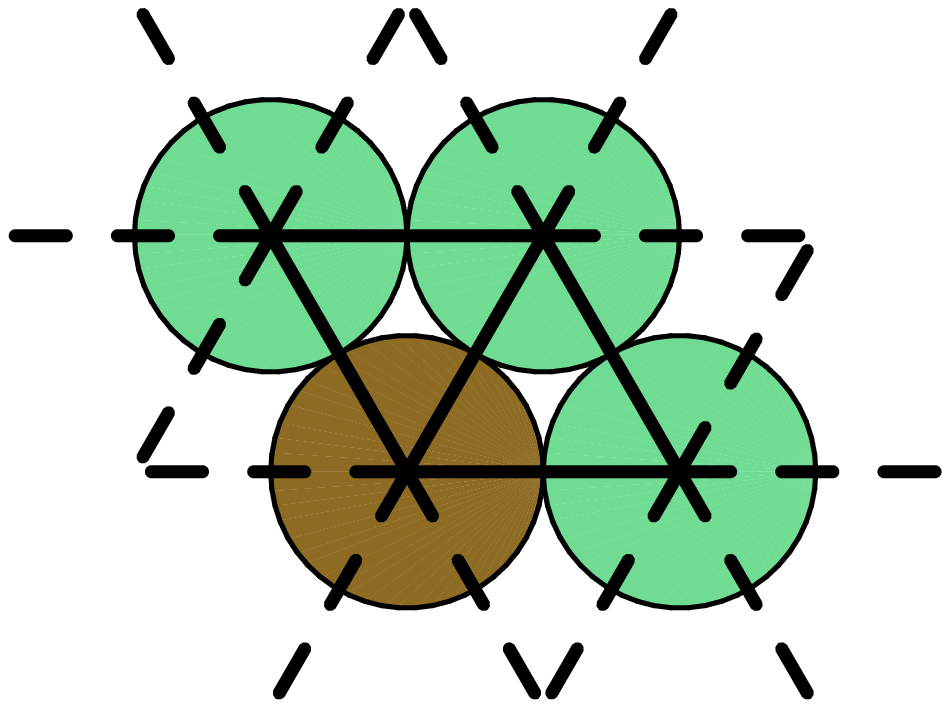}  \includegraphics[width=3.90cm]{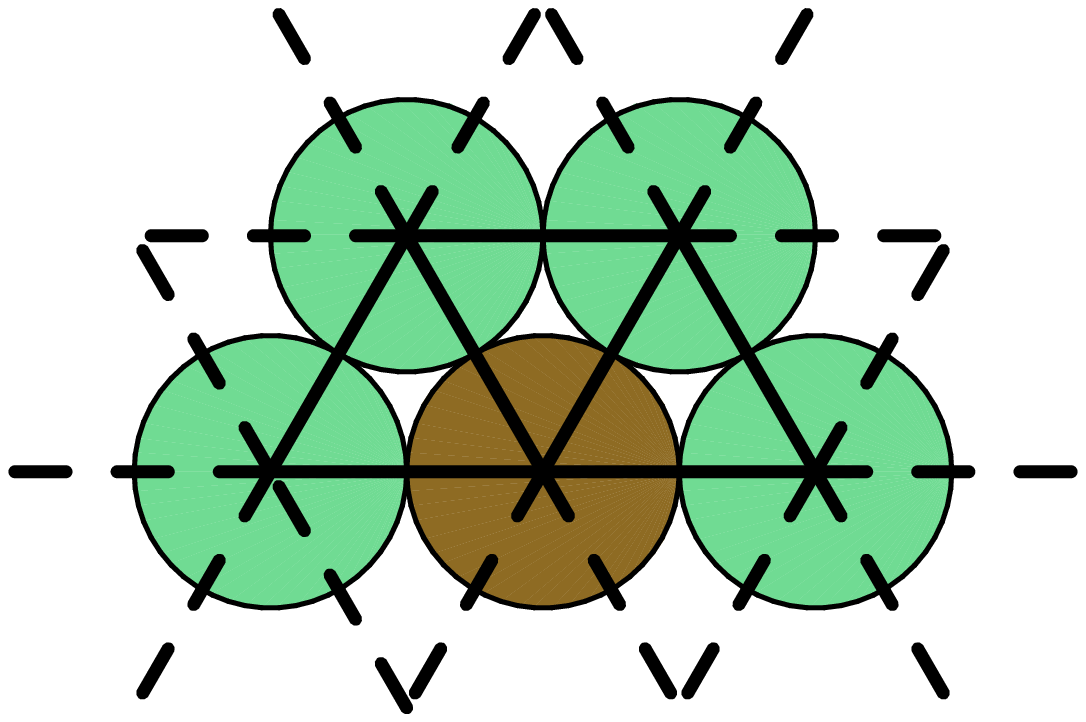}  \includegraphics[width=3.90cm]{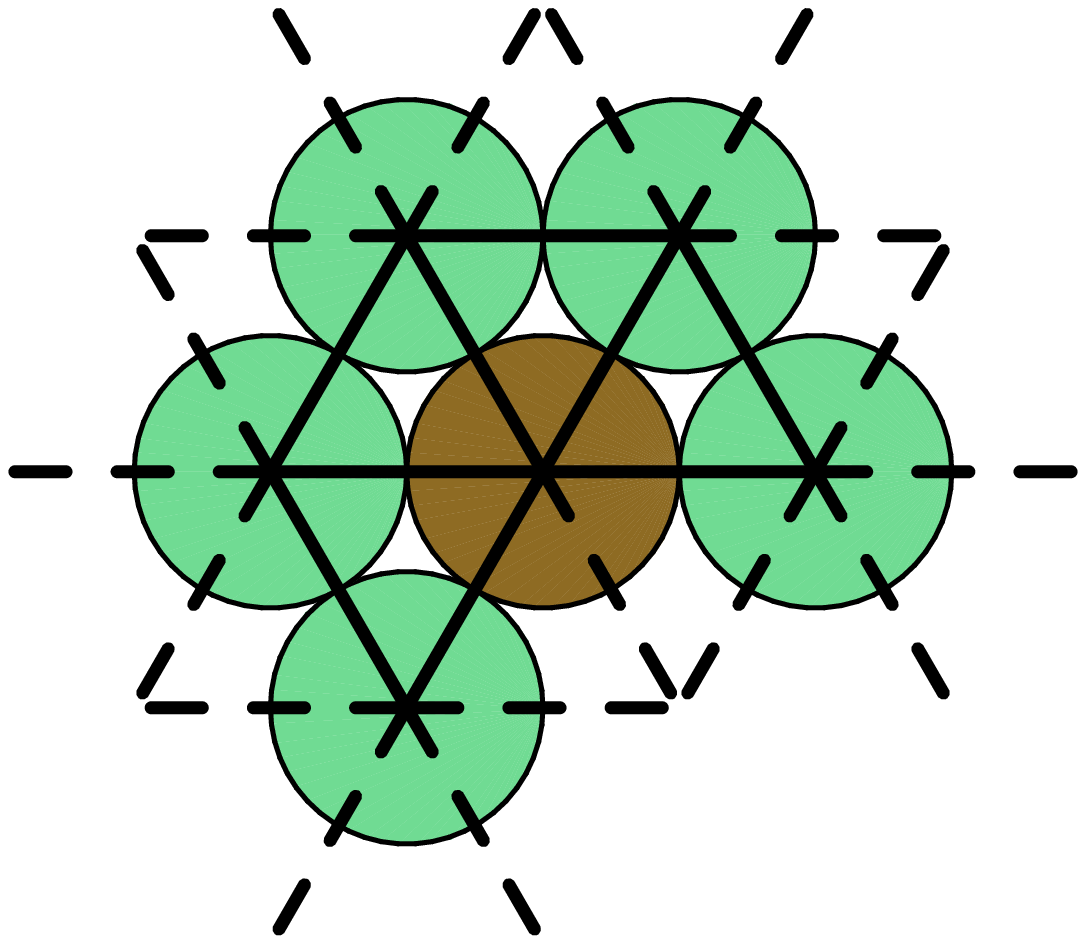}\includegraphics[width=3.90cm]{shell2.eps}
\end{center}
\caption{Optimal configurations for $N=M(1)=1,\ldots,M(2)=7$.}
\label{f.2}
\end{figure}

2. It is easy to check the correctness of Eq.~(\ref{eq:coordnmb}) for $N=1,\ldots,M(2)=7$ by direct computation (Fig.~\ref{f.2}).

\begin{figure}[htbp]                        
\begin{center}
\includegraphics[width=4cm]{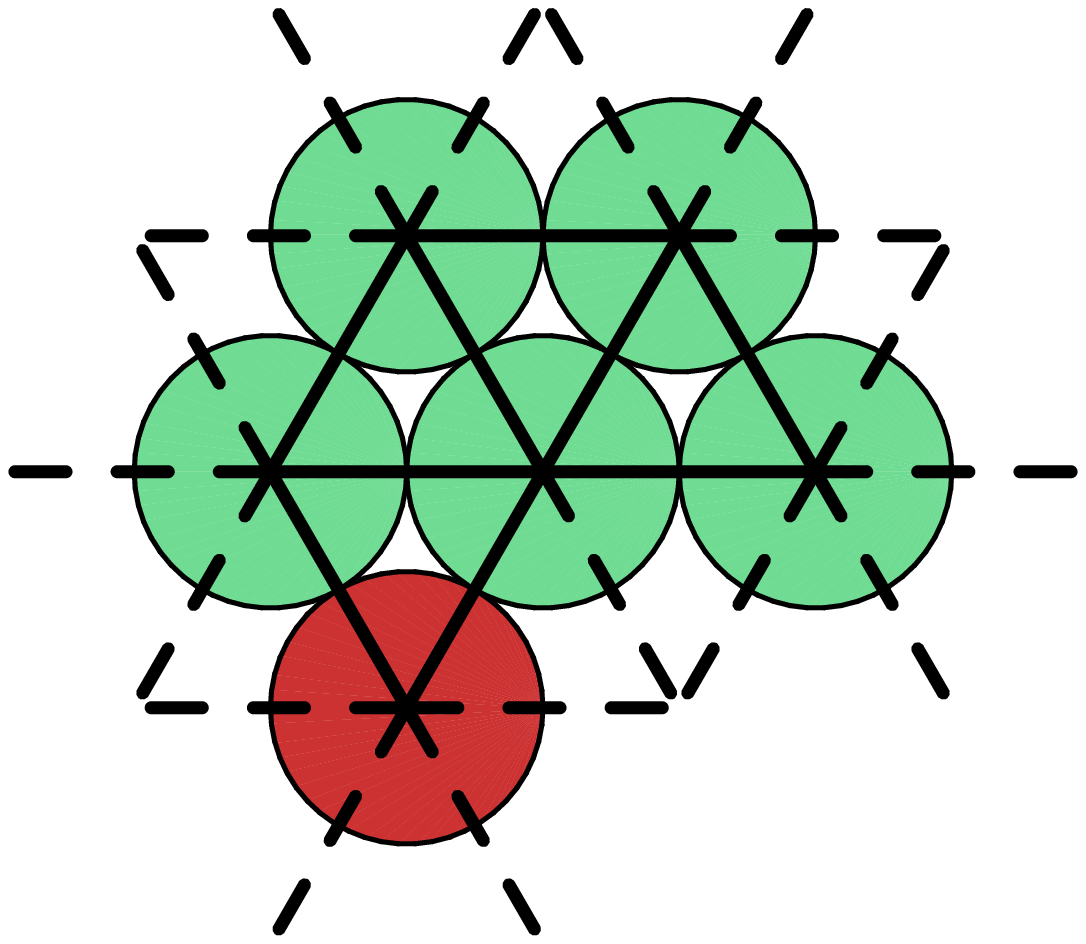} \quad \includegraphics[width=4cm]{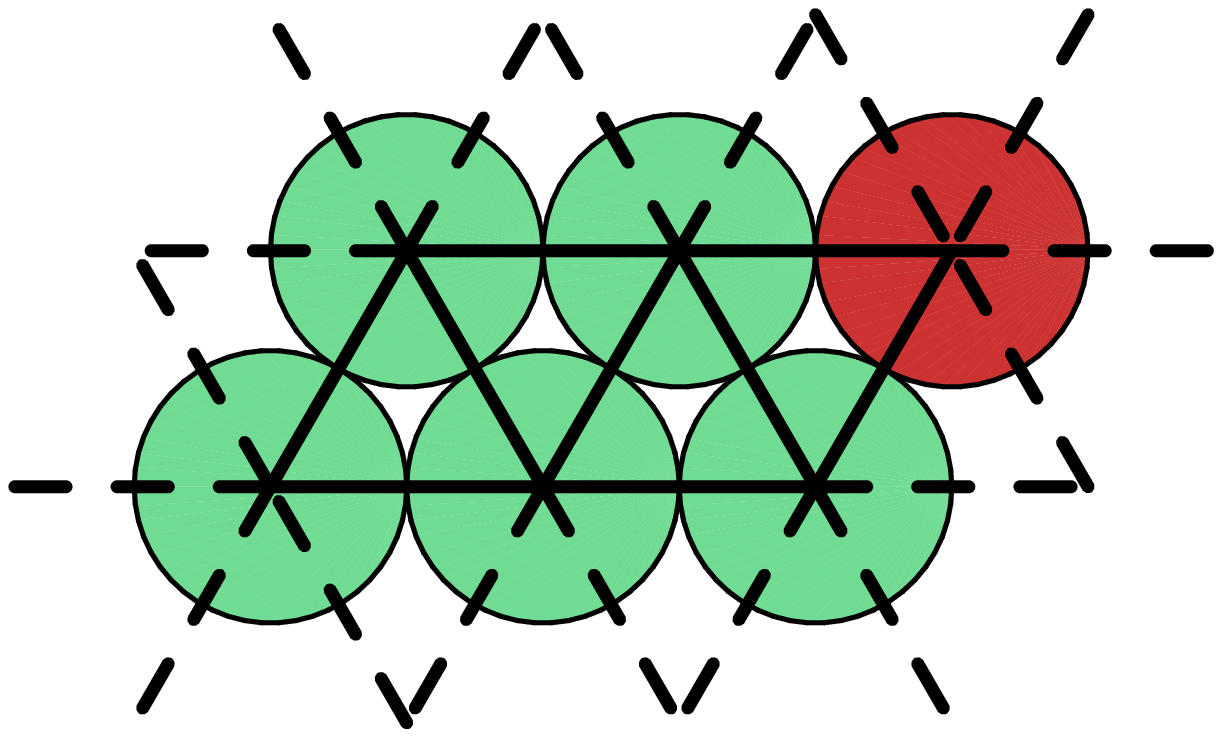} \quad \includegraphics[width=4cm]{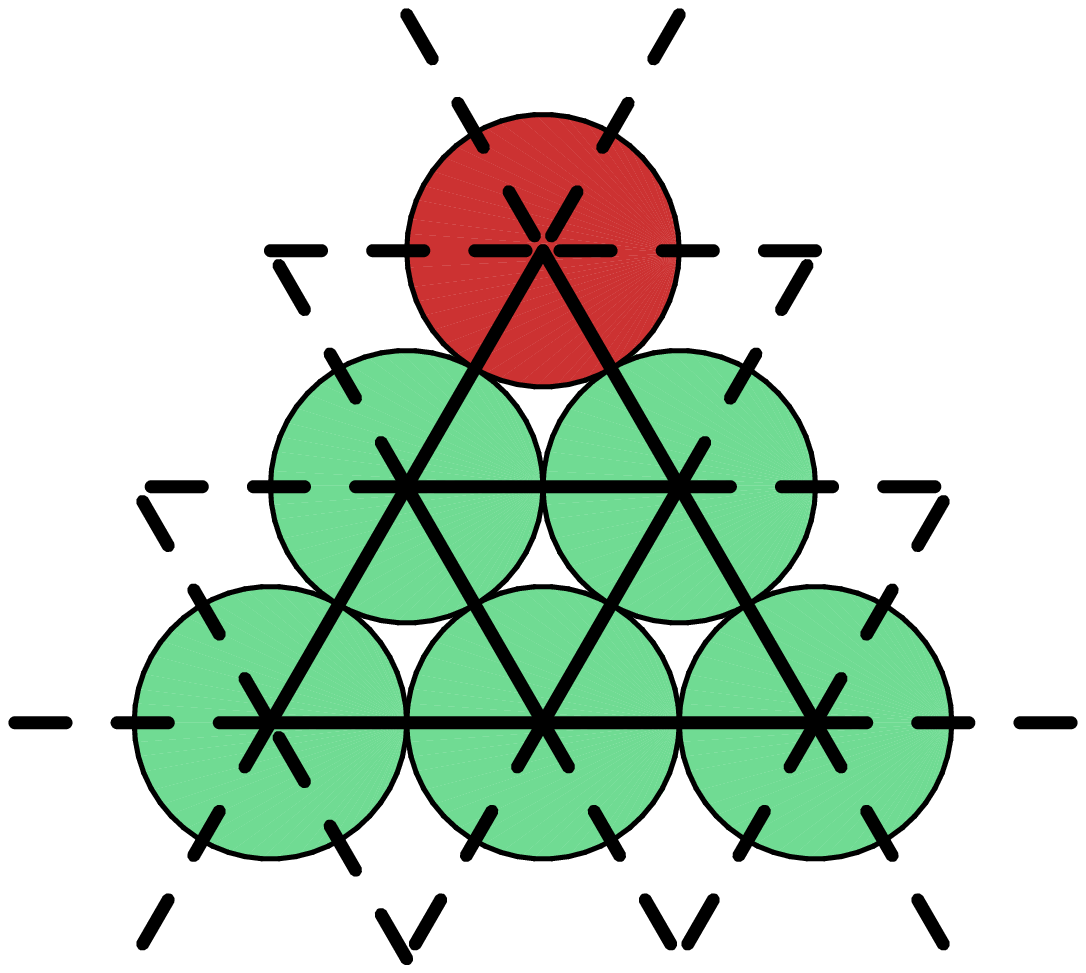}
\end{center}
\caption{All three configurations have the minimal coordination number $\alpha_6=9$.}
\label{f.3}
\end{figure}

Note that, as shown in Fig.~\ref{f.3}, the configuration with minimal coordination number is obliged to be neither unique nor convex. (Convexity may be defined as absence of cells with only one free bond.)

3. We show that Eq.~(\ref{eq:coordnmb}) is valid for $N=M(n),M(n)+1,\ldots,M(n+1)$ for any generation number $n\ge 2$
for a chosen algorithm of adding cells.

\begin{figure}[htbp]                        
\begin{center}
\includegraphics[width=3.90cm]{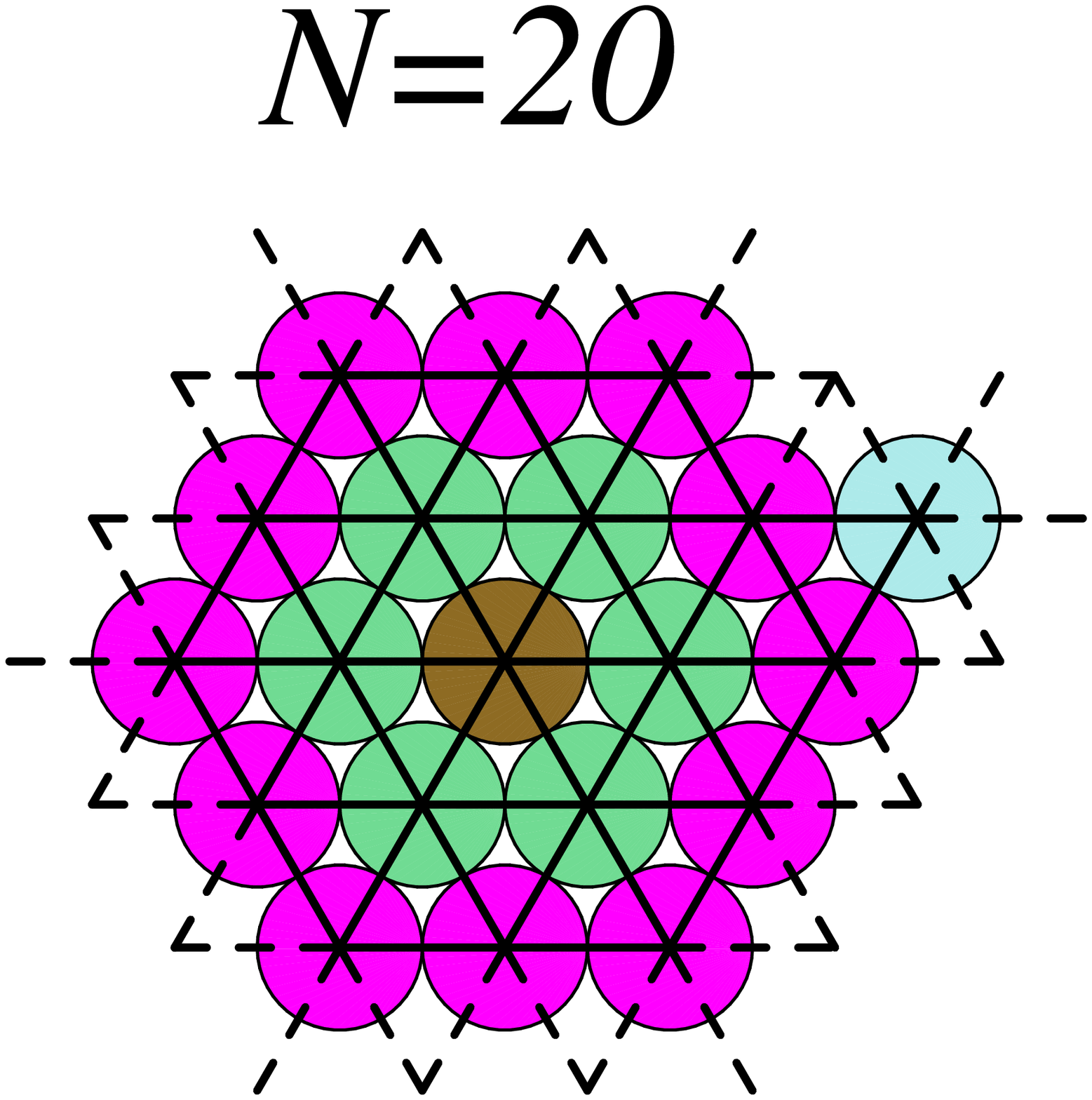}
\includegraphics[width=3.90cm]{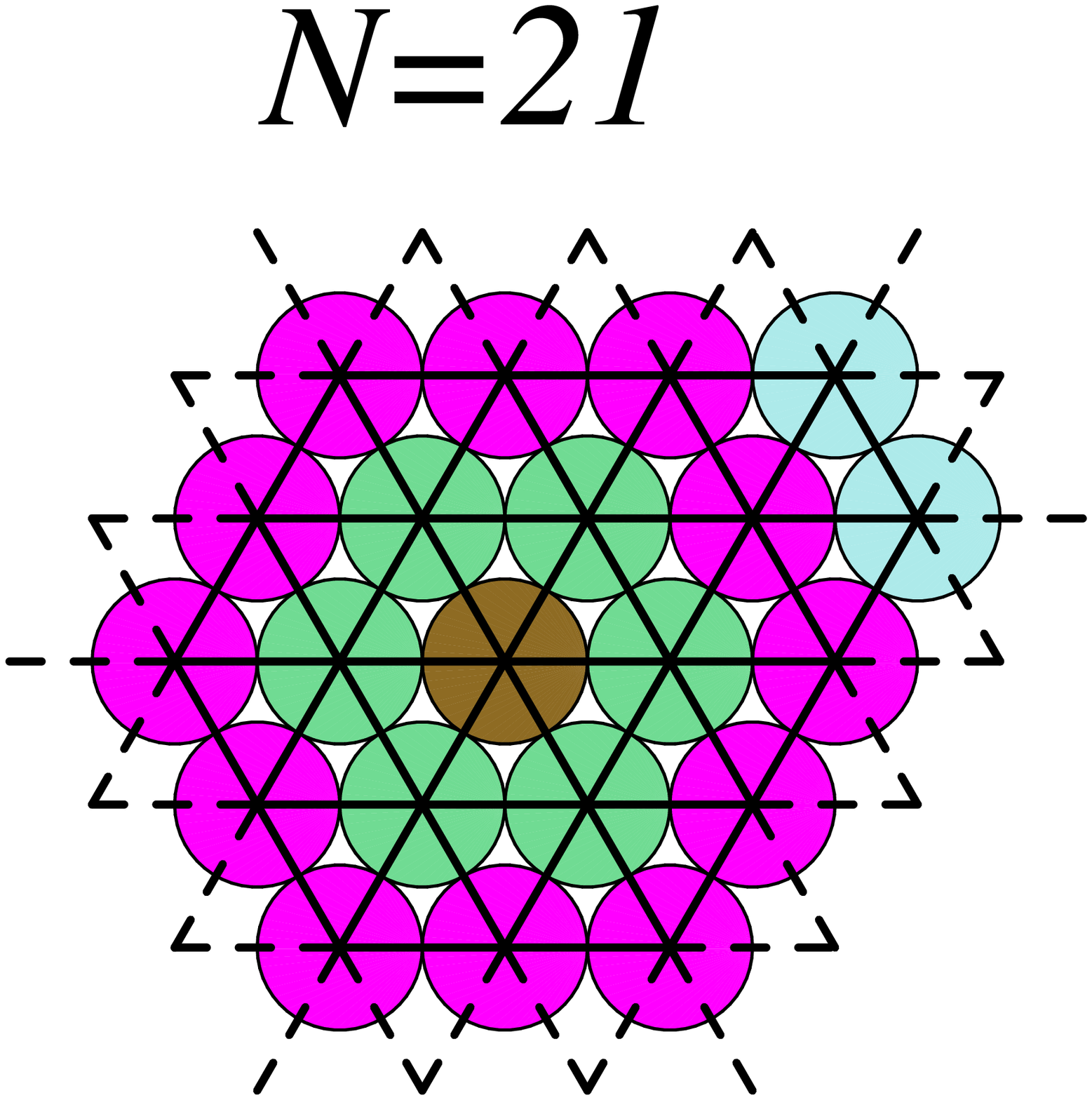}
\includegraphics[width=3.90cm]{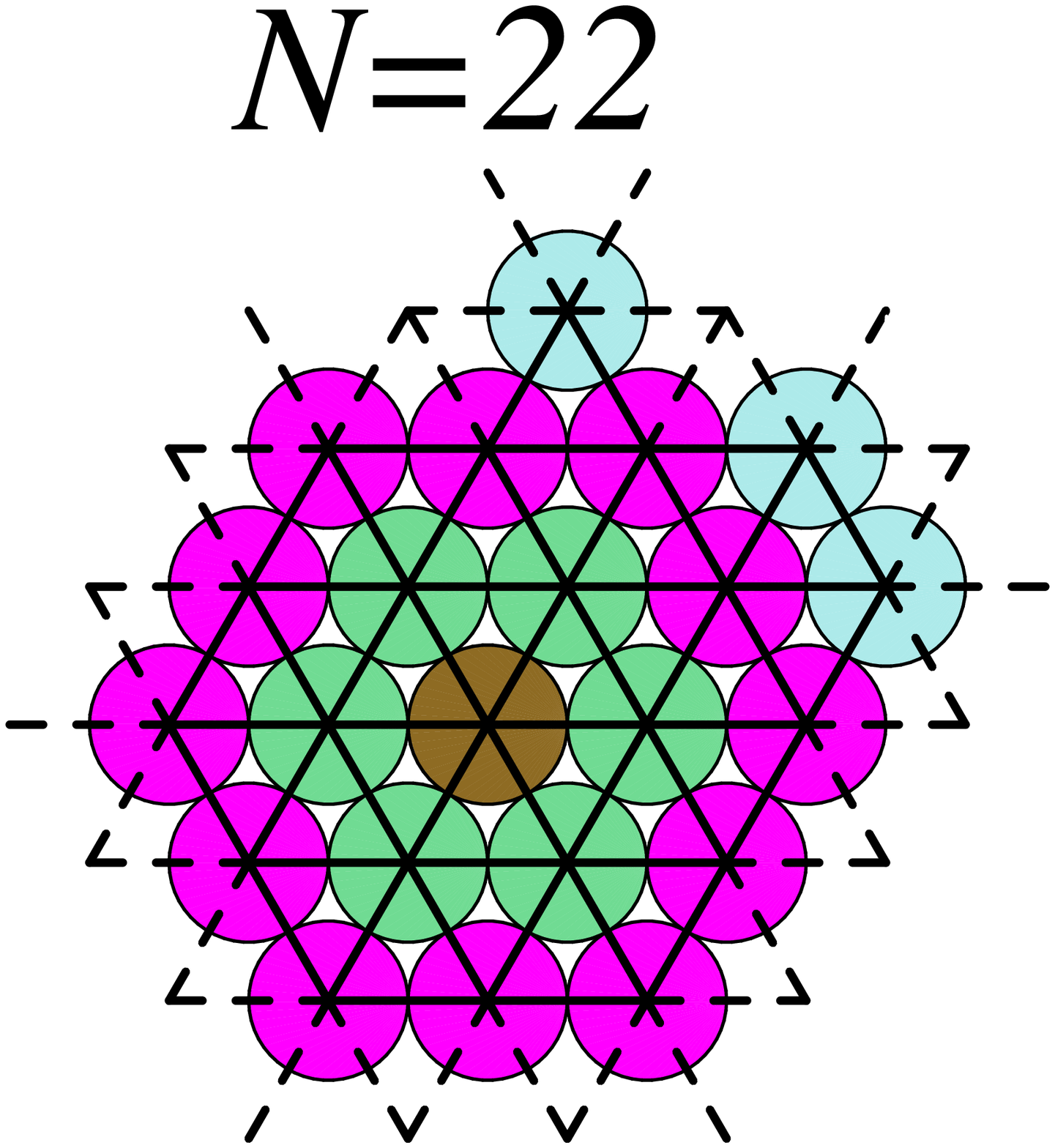}
\includegraphics[width=3.90cm]{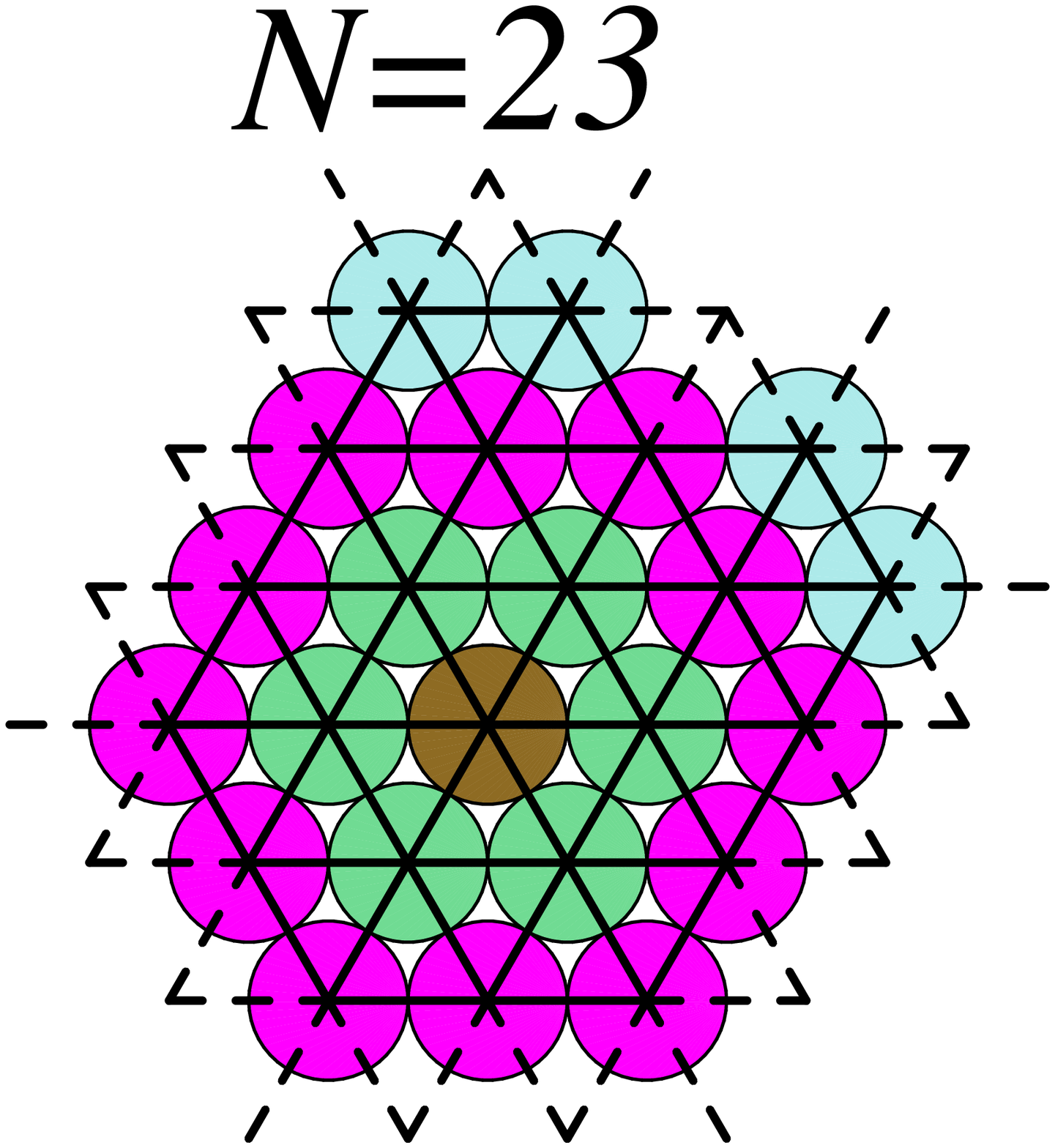}
\includegraphics[width=3.90cm]{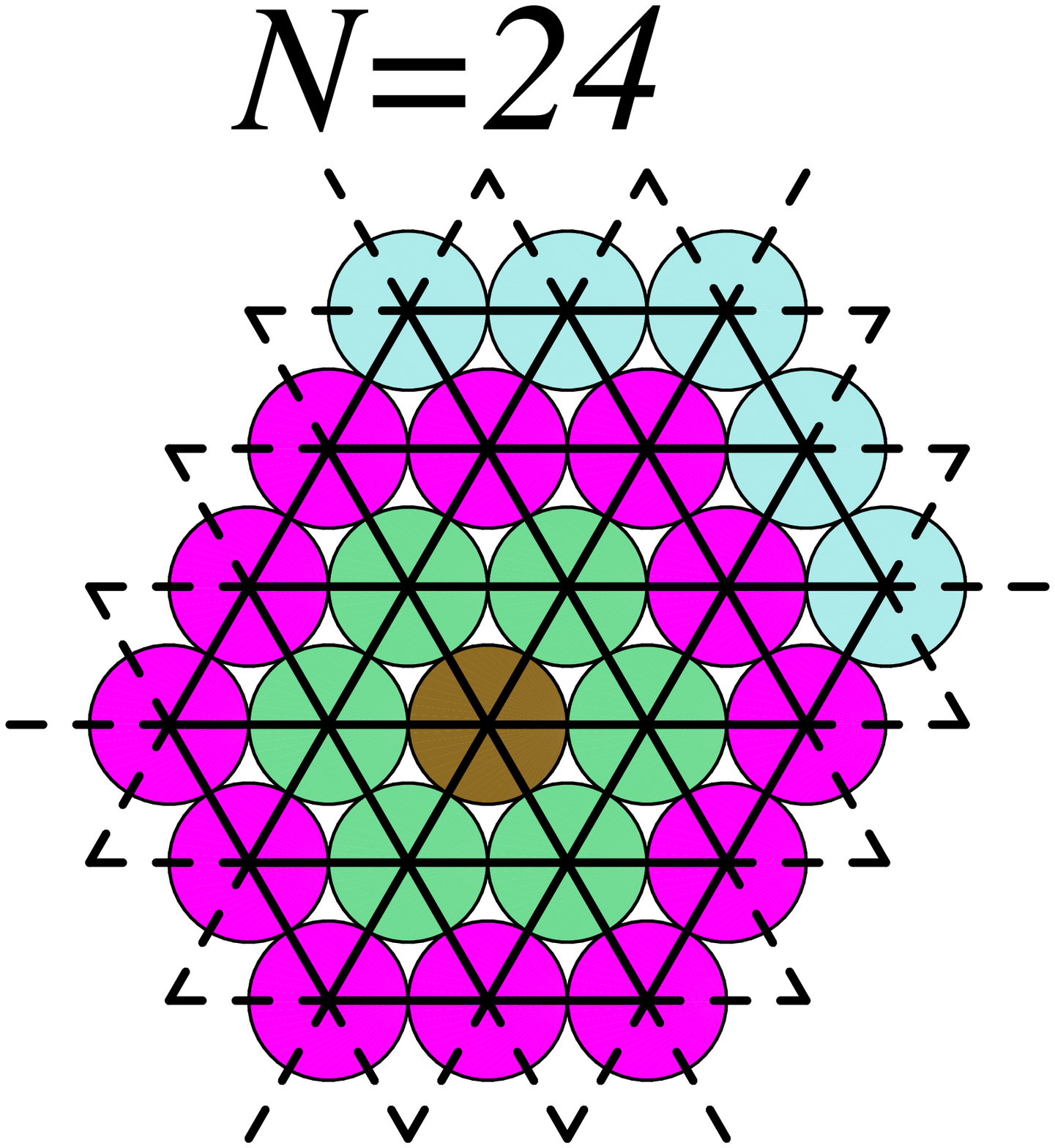}
\includegraphics[width=3.90cm]{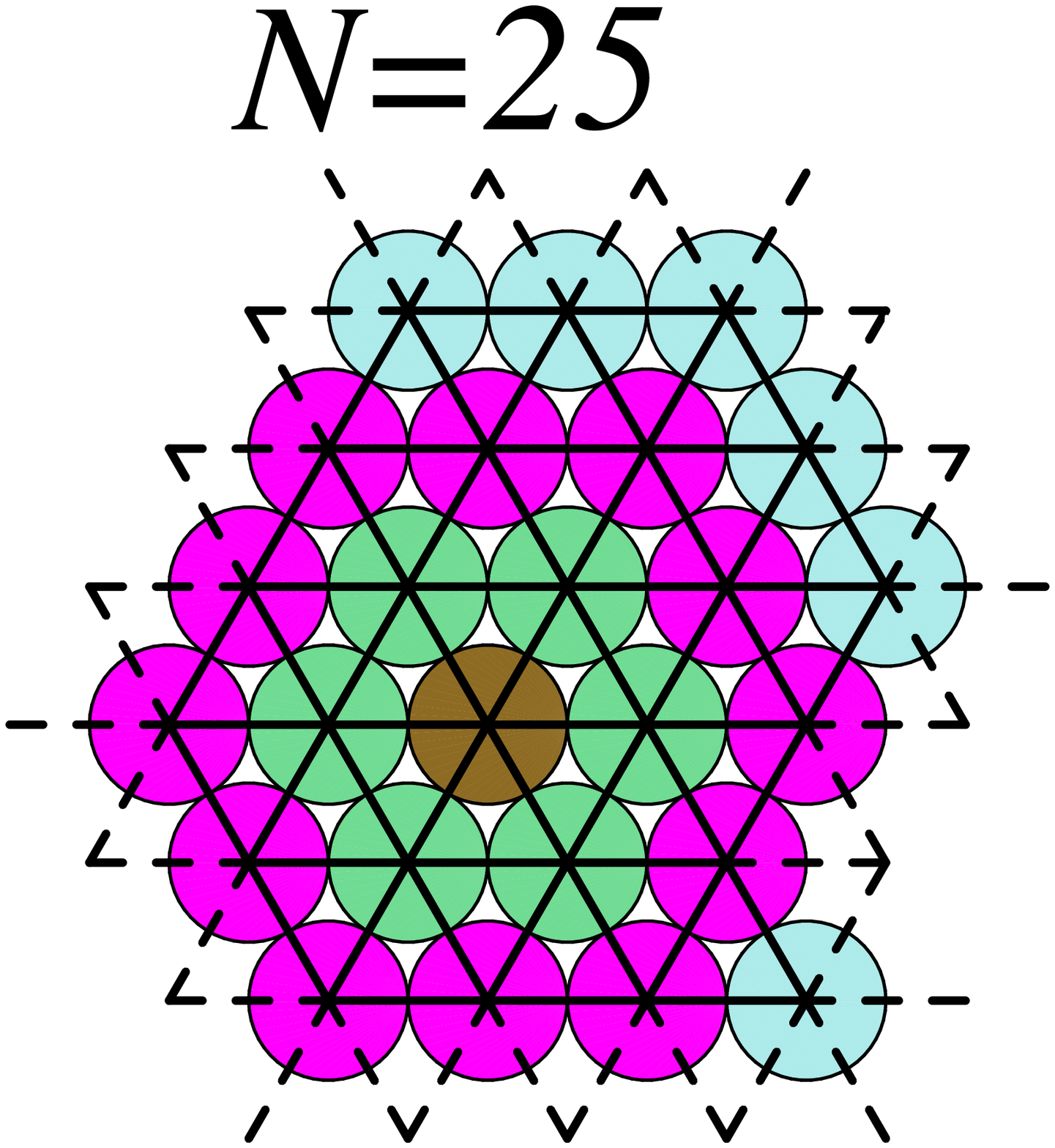}
\includegraphics[width=3.90cm]{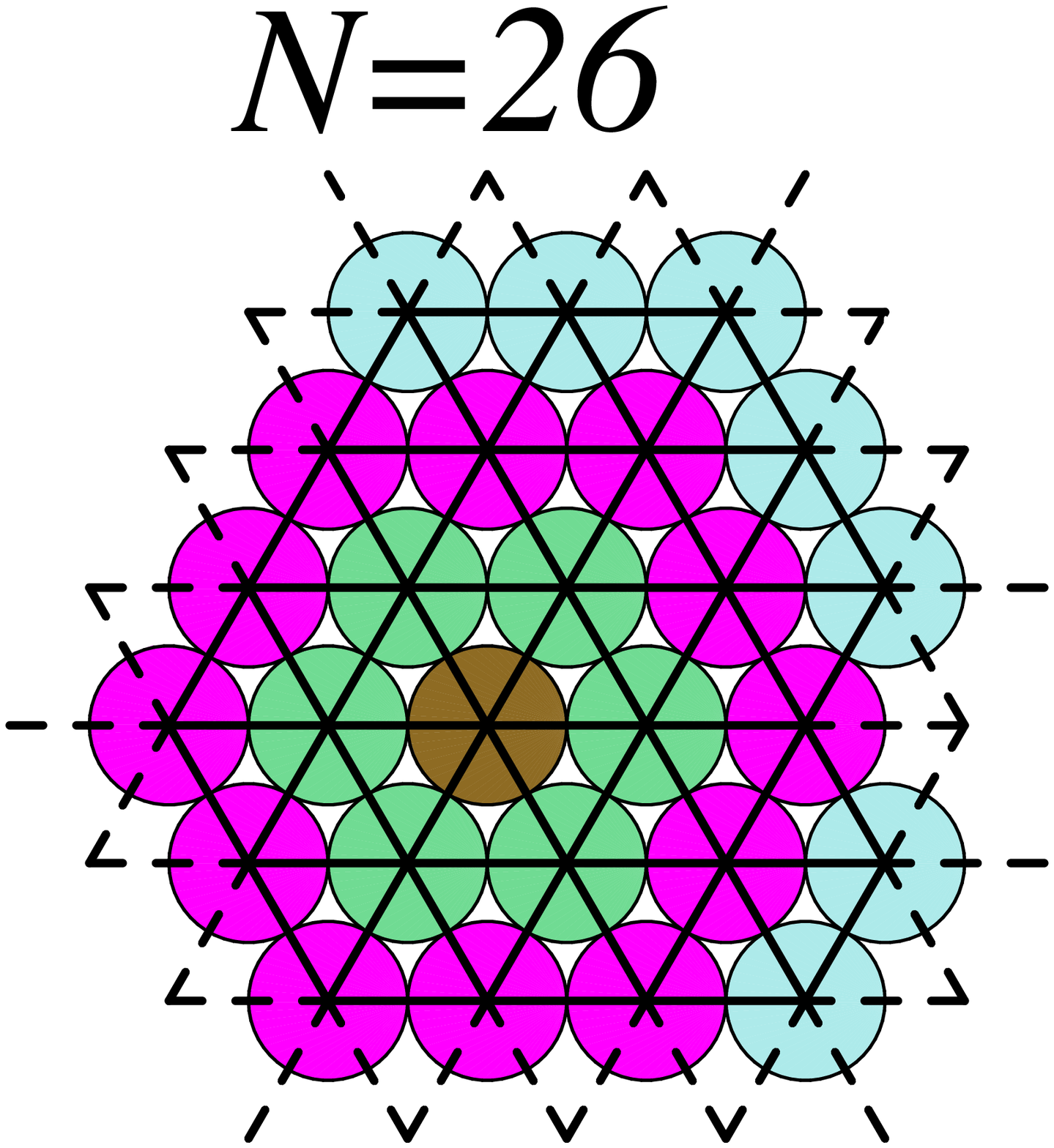}\\
\includegraphics[width=3.90cm]{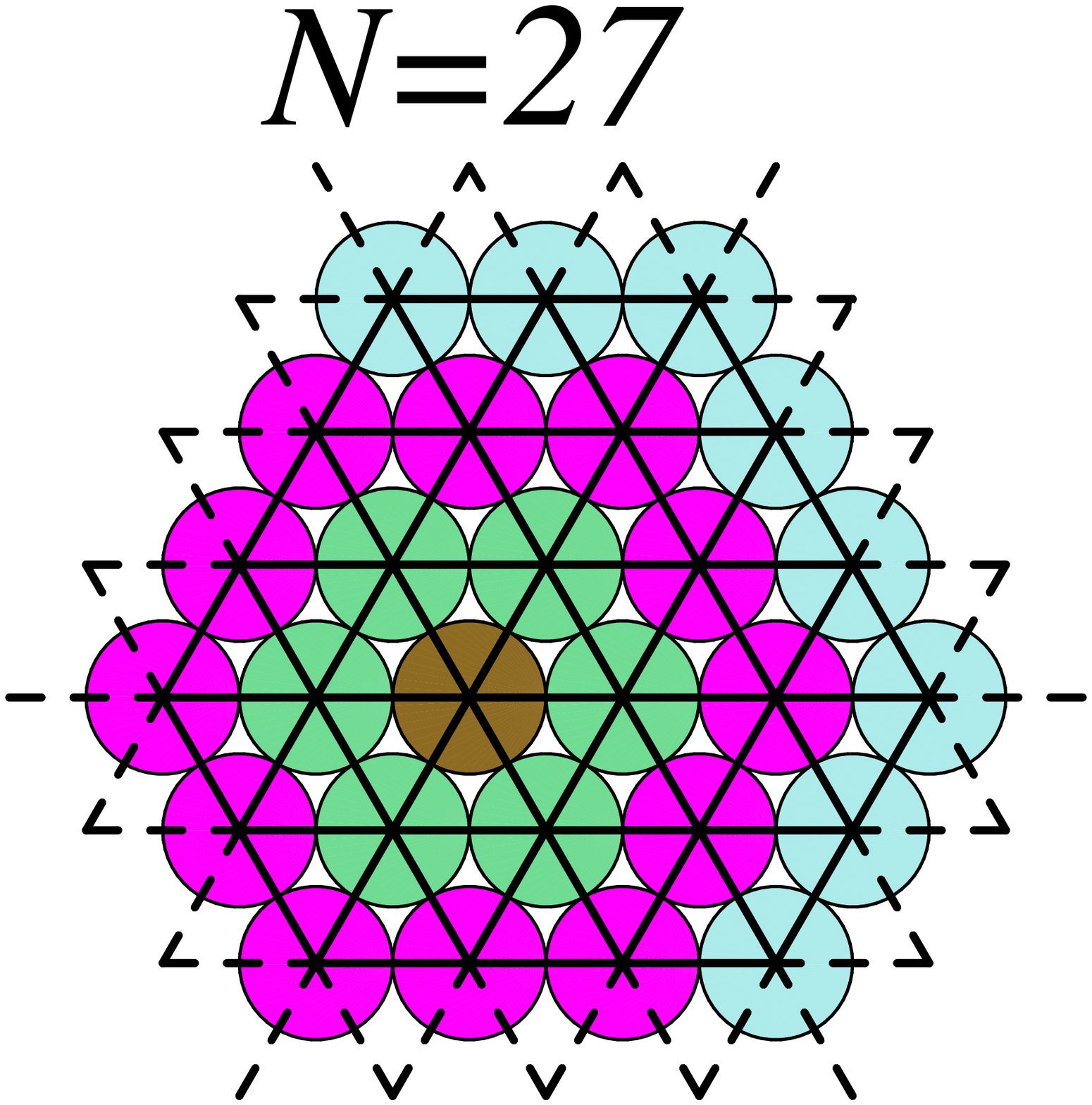}
\includegraphics[width=3.90cm]{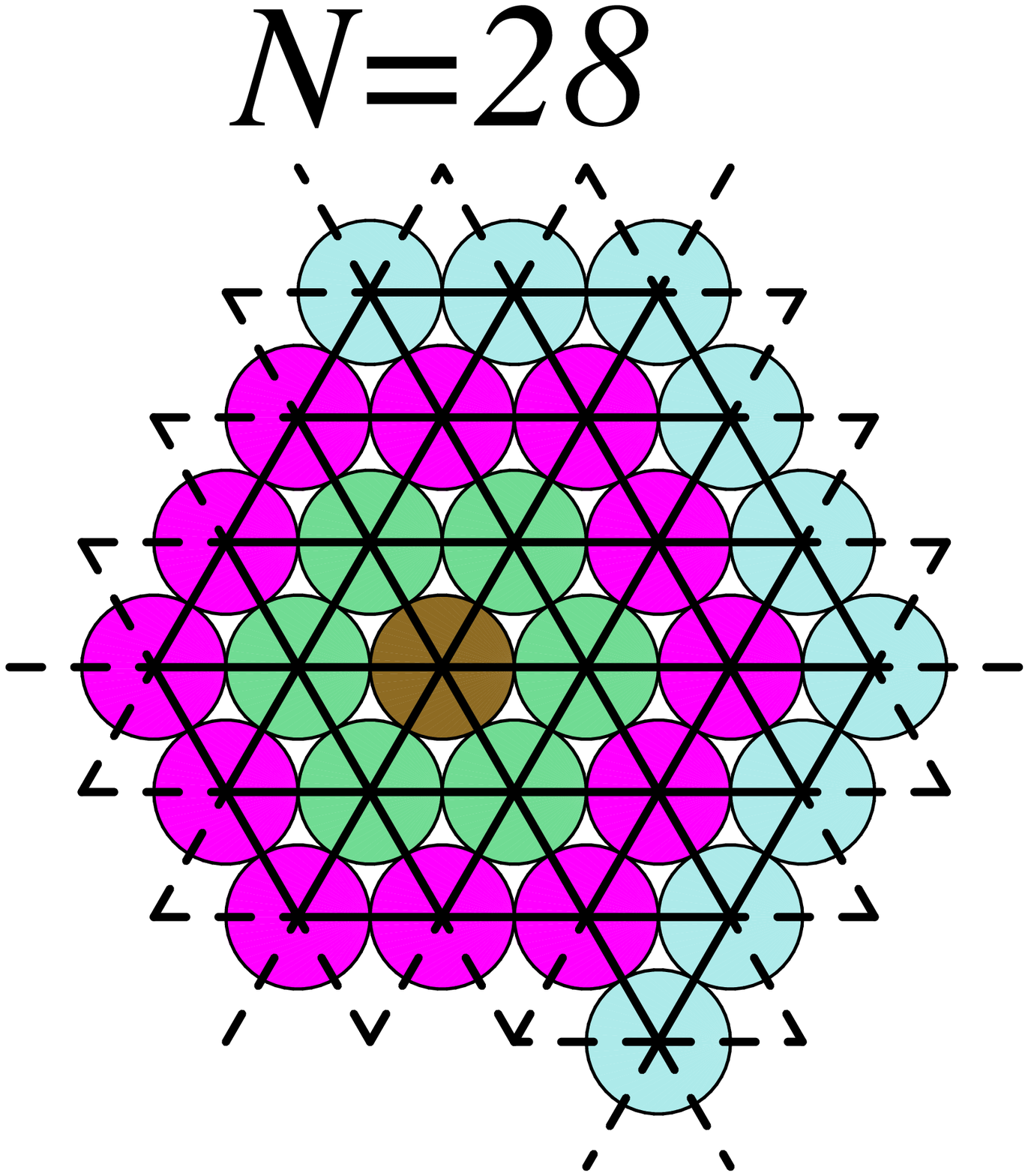}
\includegraphics[width=3.90cm]{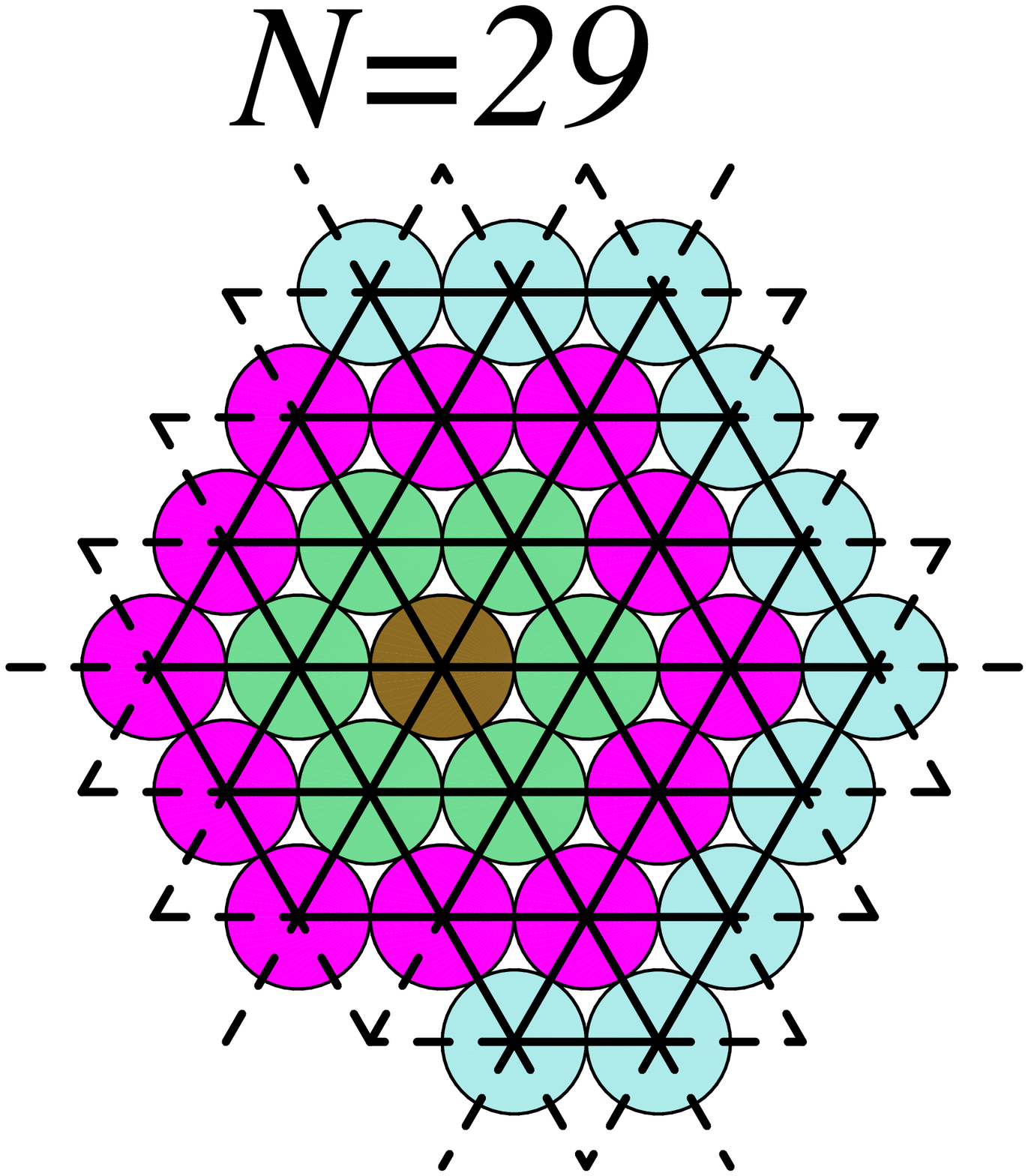}
\includegraphics[width=3.90cm]{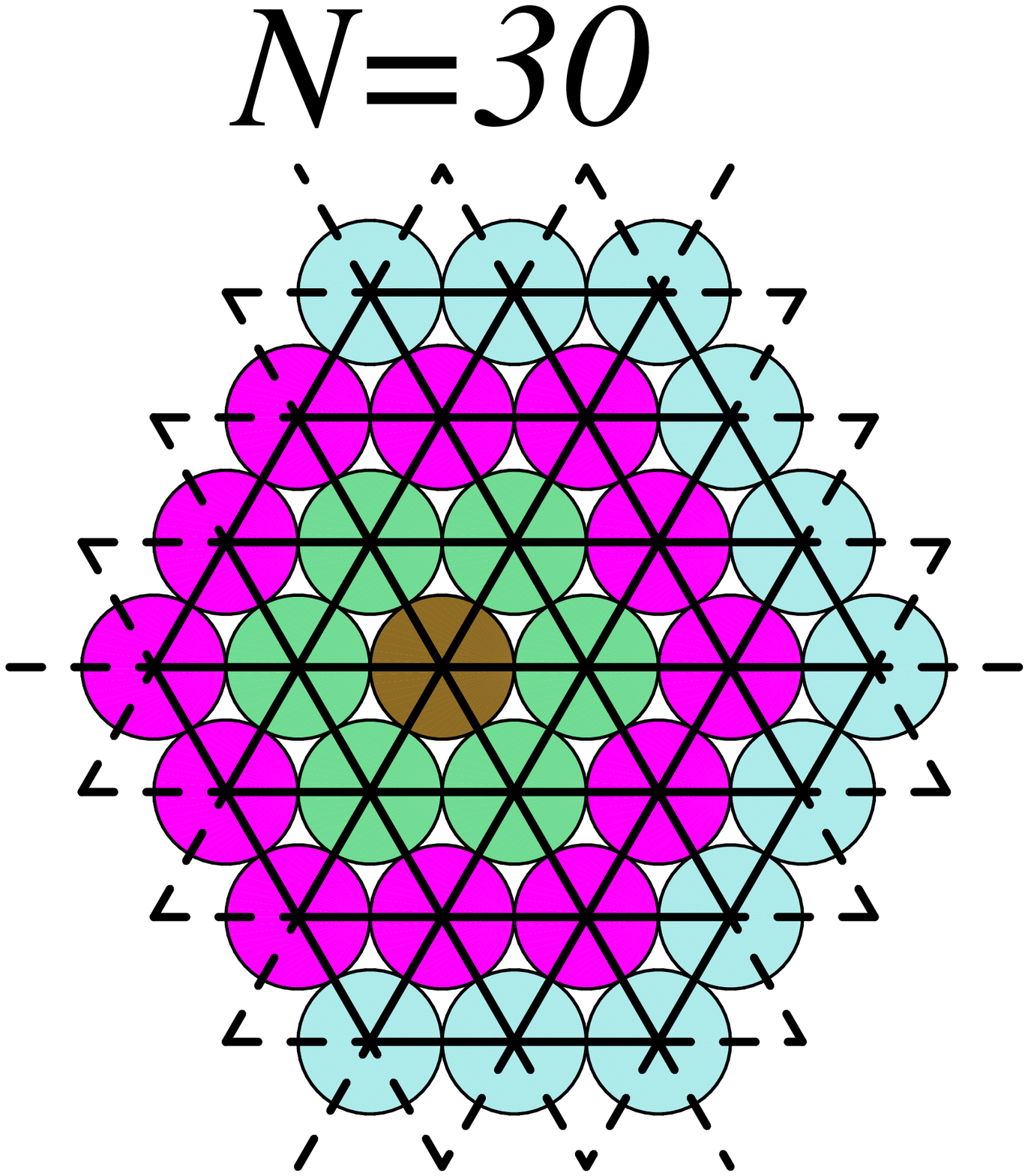}
\end{center}
\caption{Adding more cells to the complete shell of generation $n=3$. A choice of optimal sequence for $N=20,\ldots,30$.}
\label{f.4}
\end{figure}

Consider what happens when we add more cells to the complete shell of generation $n$ ($n=3$ in Fig.~\ref{f.4}). The first added is bound to two other cells independently of its place ($N=20$). So, it closes 2 existing open bonds and adds 4 new which means that the coordination number 
increases by 1: $\alpha_{M+1}=\alpha_M+\frac{1}{2}(4-2)=\alpha_M+1$. To minimize the number of exposed sides, the next cell has to be added adjacent to the first one so that it contacts with three cells ($N=21$). Since the number of new open bonds
is 3 as well, $\alpha_{M+2}=\alpha_{M+1}$. 

We can keep adding new cells until the entire side of the hexagon is filled up.
As it happens, the coordination number increases by 1, because the $(M(n)+n)$-th cell can only be bound with two others ($N=22$). The cell has to be added to one of the remaining 5 sides of 
the initial complete shell. One of the neighbouring sides has to be chosen to make 
the next increase of the coordination number as late as possible ($N=23$).
As the neighbouring side is being filled, the coordination number does not change until the entire row is completed ($N=24$).
The cycle repeats six times until all six sides of the $(n+1)$-th layer are filled up (only filling two more sides is shown
in Fig.~\ref{f.4}). Each time as the $(M(n)+i n)$-th cell takes its place ($i=1,2,3,4,5$), the coordination number adds 1.
There is no need for rearrangement of cells as more cells being added. 
Summing up, we can say that the coordination number increases each time when a cell is added to a convex configuration.

Let us now check whether Eq.~(\ref{eq:coordnmb}) follows precisely this pattern.
As seen from the equation, $\alpha_{N} = \alpha_{N-1} + \Delta$, for integers $N>1$ and $\Delta>0$, if and only if there exists an integer $K$ such that $12(N-1)-3 \le K^2 < 12N-3$. Now check what are $N$'s for which $\Delta >1$.  
The latter would imply $\sqrt{12N-3}-\sqrt{12(N-1)-3}>1$ which only possible for $N=2,3$.
Since we are now interested in the case $N \ge M(2)$, we are safe to consider only $\Delta =1$.
The changes in $\alpha_N$ may be indexed with $K$.

There are 6 changes in $\alpha_N$ between the shells of two consecutive generations, i.e. in the range of numbers between $M(n)$ and $M(n+1)$.
Showing first that $\alpha_{M(n)+1}= \alpha_{M(n)} +1$, let us set $K=\alpha_{M(n)}=6n-3$. Then $12M(n)-3\le (6n-3)^2 < 12(M(n)+1)-3$.
Indeed, $12(3n^2-3n+1)-3 = (6n-3)^2 < 12(3n^2-3n+2)-3$.

As explained above,
the remaining 5 increases of $\alpha_N$ should take place for $N=M(n)+i n$, $i=1,2,3,4,5$.
To prove this, take $K=\alpha_{M(n)} + i$, then we can write
\begin{eqnarray}
12(M(n)+i n -1)-3 < (\alpha_{M(n)} + i)^2 
< 12(M(n)+i n) -3, \nonumber \\
12(3n^2-3n+1+i n -1)-3 < (6n-3+i)^2 
< 12(3n^2-3n+1+i n)-3, \nonumber \\
36n^2+12(i-3)n-3 < 36n^2+12(i-3)n+(i-3)^2 
< 36n^2+12(i-3)n+9, \nonumber
\end{eqnarray}
which is obviously satisfied. Therefore, $\alpha_{M(n)+i n} = \alpha_{M(n)+i n -1} +1$.

Thus we see that, according to our formula, $\alpha_N$ jumps by 1 
when the first cell is added to the complete shell and when
a portion of $n$ cells is added to the 
complete shell of generation $n$ unless the $n+1$ layer is completed. 
By way of induction we conclude that Eq.~(\ref{eq:coordnmb}) is valid for all $N \ge 1$.

4. The remaining thing to do is to show that the above scenario is optimal in the sense that it provides the minimal number of exposed sites for every number of cells.
Clearly, an optimal configuration must be simply connected. 
Let us take an arbitrary simply conected cluster of $N$ cells.
There exists a simple closed polygonal line that connects the centres of all $p$ cells with exposed bond sites (we call it a perimeter).
Let us orient the perimeter so that the interior of the cluster ($N-p$ cells) be on the left.

Each perimeter cell can belong to only one of the following 4 groups. The first group consists of $t$ cells where the perimeter line makes no turn.
Such a cell has two free bonds on the right and at most two bonds with the interior part of the cluster. The second group includes $u$ cells where the perimeter turns
to the left through $\pi/3$. The members of the second group have three exposed bond sites on the right and at most one bond with the interior.
The third group is made up with $v$ cells with the $2\pi/3$ turns of the perimeter and with 4 free bonds on the right. The cells of this group are bonded 
only with two perimeter neighbours. The last fourth group is formed with $w$ cells in which the perimeter line turns to the right through $\pi/3$.
Such perimeter cells cannot appear in a convex cluster. They have one free bond on the right and up to three bonds to the interior. Thus, we have
\begin{equation}
p=t+u+v+w.
\label{eq:p}
\end{equation}

When traversing the complete path along the perimeter, we make exactly one turn (the winding number of the perimeter is one), hence we can write
\begin{equation}
u+2v-w=6.
\label{eq:wind}
\end{equation}

By definition, the coordination number of the $N$-cluster $\alpha_N$ equals half the number of exposed bond sites on the right of the perimeter.
Summing up these exposed sites gives 
\begin{equation}
2\alpha_N=2t+3u+4v+w.
\label{eq:alpha}
\end{equation}

Eliminating the quantities $t,u,v,w$ from Eq.~(\ref{eq:alpha}) with the help of Eqs.~(\ref{eq:p}), (\ref{eq:wind}) leads to
\begin{equation}
\alpha_N=p+3 .
\label{eq:alphaN}
\end{equation}
The last equation simply shows that the coordination number is essentially the length of the perimeter, i.e. the number of the exposed cells.

All the bonds with the interior contribute to the coordination number of a cluster that is obtained after peeling off the $N$-cluster. Then, the coordination number of this ($N-p$)-cluster can be estimated as
\begin{equation}
2\alpha_{N-p} \le 2t+u+3w.
\label{eq:alphaNp}
\end{equation}
Again, with the help of Eqs.~(\ref{eq:p}), (\ref{eq:wind}) we obtain from Eq.~(\ref{eq:alphaNp})
$$ 
\alpha_{N-p} \le p-3.
$$ 
Combining the last inequality with Eq.~(\ref{eq:alphaN}) allows us to relate the coordination numbers of the full and the stripped clusters
\begin{equation}
\alpha_N \ge \alpha_{N-p} + 6.
\label{eq:alphaNN}
\end{equation}

Let us now assume that the coordination number of the peeled-off $(N-p)$-cluster satisfies Eq.~(\ref{eq:coordnmb})
for the minimal coordination number. Then we can write
$$ 
\alpha_{N-p} \ge \sqrt{12(N-p)-3} = \sqrt{12(N-\alpha_N)-33}
$$ 
(here we used Eq.~(\ref{eq:alphaN}) to eliminate $p$ in the right-hand side).
Substitution of $\alpha_{N-p}$ from the last inequality into Eq.~(\ref{eq:alphaNN}) and some algebraic transformations
eventually lead to the required estimate for the $N$-cluster
$$ 
\alpha_{N} \ge \sqrt{12N-3} .
$$ 

Thus we see that it is impossible to lower the coordination number below the value that was 
computed for the sequence of clusters we considered above in the former parts of this proof.
We conclude that indeed such a scenario of growth guarantees the most compact configuration for any number of cells with the minimal
coordination number from Eq.~(\ref{eq:coordnmb}).

Proof of Eq.~(\ref{eq:coordnmb_sq}) is similar.

\section{Discussion}

1. The minimal coordination number formula may be used to compute the number of actual bonds
$U(N)=3N-\left\lceil\sqrt{12 N - 3}\right\rceil$
(for hexagonal lattice).
For complete hexagonal shells with $N=M(n),  n = 1,2,\ldots$ (Fig.~\ref{f.1}), the ceiling function can be omitted.
As suggested by Ishimoto and Kikuchi\cite{Ishimoto08}, the expression $V(N) = 3N-\sqrt{12 N - 3}$ can be used as an approximation for arbitrary $N$.
They present a comparison of $V(N)$ for real $N$ with what they call ``the exactly counted discrete function''
(though they plot two {\it continuous} functions). It is not clear from the paper how ``the exactly counted discrete function'' was computed.
One can see from the graph (Fig.2 in \cite{Ishimoto08}) that
the latter function differs from $U(N)$ (e.g., notice too big difference between the curves for $N=30$).
Anyway, our theory provides a solid proof that $V(N)$ can be used as an approximate function for large $N$ since it cannot differ from
the exact optimal function by more than 1.

2. Equation~(\ref{eq:coordnmb}) may be ``inverted'' to give an equation for $N_\alpha$, the maximal number of strands in the bundle for
given (integer) coordination number $\alpha$:
$N_\alpha = \lfloor (\alpha^2 + 3) /12 \rfloor$, $\alpha \ge 5$,
where$\lfloor\cdot\rfloor$ stands for the floor function: for $x \in \mathbb{R}$, $\lfloor x\rfloor$ is defined as the greatest integer 
less than or equal to $x$.
For $\alpha = 9$ and $\alpha \ge 11$,
the $N_\alpha$-bundles are considered as particularly stable in a sense that
they have the same surface energy as the $(N_\alpha - 1)$-bundles: addition of the $N_\alpha$-th filament 
does not affect the coordination number. However, one more filament increases the surface energy. 
Such numbers $N_\alpha$ are called magic and supermagic~\cite{Pereira00,Schnurr02}.
For the square lattice the ``inverted'' Eq.~(\ref{eq:coordnmb_sq}) reads $N_\beta = \lfloor \beta^2 /4 \rfloor$, $\beta \ge 2$.

3. A word of warning should be placed here concerning the use of the mininal coordination number to  estimate
the surface energy of the bundle. The coordination number, we consider in this paper, is actually defined for a thin
cross-sectional slice of a bundle. To compute the surface energy of the entire bundle, one need to integrate
over its length. Since in general the lengths of the particular strands may differ, the total number of the exposed binding sites does not necessarily amount to simply a doubled product of the coordination number and the length of the bundle axis.
To get the correct result one should account for the actual shapes of individual filaments exposed to the environment
and these shapes could be rather complex. This subject is beyond the scope of this paper.

It is only worth to mention here that, for {\it closed} parallel bundles the lengths of the components are quantized and this fact simplifies the computation of the surface energy reducing it essentially to the coordination number multiplied by the length. However, the life is not so simple even in this case because there are intrinsic geometric constraints on the arrangements of the rods into the bundle which are not always consistent with the minimality of the exposed surface~\cite{Starostin06}.
It may also be the case that the sequence of optimally packed bundles is kinetically unfeasible.
For example, when the filament coils into a toroidal bundle it is unlikely that it passes through the hole.


4. Let us here focus on the hexagonal lattice first. The complete shells are indexed with the generation number $n$
and the number of cells is $N_\alpha=3n^2-3n+1, n=1,2,\ldots$. Then, for any $N$-bundle we can define a real 
quantity $\nu_\alpha(N) \in \mathbb{R}, \nu_\alpha \ge 1$ such that $N=3\nu_\alpha^2-3\nu_\alpha+1$.

The problem of finding a configuration that minimizes the coordination number may be considered as an
isoperimetric problem where $N$ plays part of the ``area content" and the coordination number is essentially
the perimeter length (see Eq.~(\ref{eq:alphaN})). In the classic continuum analogue, the solution of the 2D isoperimetric problem 
is a circle with area $A=\pi r^2$ and circumference $S=2\pi r$ so that $\frac{dA}{dr}=S$ with
the radius $r$ being the so-called ``harmonic parameter"~\cite{Fjelstad03}.
In the discrete problem, the last equation transforms into
$\left\lceil\frac{dN}{d\nu_\alpha}\right\rceil = \left\lceil 6\nu_\alpha - 3\right\rceil =\left\lceil\sqrt{12 N - 3}\right\rceil = \alpha_N$. Thus, for complete shells $\nu_\alpha \equiv n$ serves as the harmonic parameter and
the above equation may be viewed as a discrete analogue of the known expression for the perimeter of the regular polygon as a derivative of its area relative to the radius of the inscribed circle~\cite{Emert97}.
Note that the latter relationship follows from the Steiner-Minkowski theorem~\cite{Berger87b}.
Thus, parallels can be seen between the sequence of the solutions to the discrete isoperimetric problem and the painting (or collaring) of polygons.

For the square lattice, we have correspondingly $\nu_\beta(N)=\sqrt{N}$ and $\left\lceil\frac{dN}{d\nu_\beta}\right\rceil = \beta_N$.

5. The proven extremality property of layered packing may look a little counterintuitive for thick bundles with a large number of rods.
Indeed, one may expect that the optimal shape of the cross section would approach a circle as much as possible.
As we can prove now, this is not the case~\cite{Schnurr02}. 
Moreover, the relative difference of the minimal shape and the circle becomes more noticeable as $N$ grows.
To see this let us take a complete hexagon with $M(n)$ cells. Its perimeter is a regular hexagon and it may be inscribed into a circle of radius $R=(n-1)d$, where $d$ is the diameter of an individual rod.
The maximal distance between the circle and the inscribed hexagon is proportional to the radius and equals $(1-\sqrt{3}/2)R$.
Each of six circular segments appeared between the hexagon and the circle will make a room for, say, a couple of additional layers if $n \gtrsim 16$ ($M(16)=721$).
This suggests that these gaps could be filled in with two or more layers as more rods join the bundle.
On the other hand, the optimal scenario of growth prescribes that a layer will be filled completely until the next regular hexagon with $M(n+1)$
is formed.
Of course, as we have seen, the optimal configurations may not be unique (Fig.~\ref{f.3}) and
some of them could be more ``circular'' than others,
but diversions to them in the process of growth would imply rearrangements of rods which could be prohibited.


\newpage


\bibliography{../../star}
\bibliographystyle{plain}

\end{document}